\documentclass[prb,twocolumn,superscriptaddress]{revtex4-2} 
\usepackage{mathrsfs}
\usepackage{amsfonts}
\usepackage{amsmath}
\usepackage{txfonts}
\usepackage{amssymb}
\usepackage{graphicx,subfigure}
\usepackage{bm}
\usepackage{color}
\usepackage[normalem]{ulem}
\usepackage{dcolumn} 
\usepackage{bbold}
\usepackage[table]{xcolor}
\usepackage{float}
\usepackage{tabularx}
\usepackage{hyperref}
\newcommand{\ket}[1]{|#1\rangle}
\newcommand{\bra}[1]{\langle #1|}

\begin{document}

\title{Weighted Berry curvature and global geometry of mixed quantum states} 

\author{Dominik Kuczy\'nski}
\affiliation{Department of Physics and Astronomy, Uppsala University, Box 524, 
SE-751 20 Uppsala, Sweden}

\author{Erik Sj\"oqvist}
\affiliation{Department of Physics and Astronomy, Uppsala University, Box 524, 
SE-751 20 Uppsala, Sweden}
\email{erik.sjoqvist@physics.uu.se}

\date{\today}

\begin{abstract}
We examine the geometric interpretation of the quantum geometric tensor proposed in 
[Phys. Rev. B {\bf 110}, 035404 (2024)] for mixed quantum states, focusing on its imaginary 
part, which is proportional to a weighted sum of the Berry curvatures of the eigenstates of 
the density operator. While the real part naturally decomposes into Fisher--Rao and weighted 
Fubini--Study contributions, we show that the imaginary part does not, in general, coincide 
with the curvature of a connection on a globally defined U(1) line bundle whose holonomy 
yields a mixed-state geometric phase. Using a two-level system as an explicit example, we 
demonstrate that its surface integral depends on the choice of surface bounded by the same 
closed path, with ambiguities that are not integer multiples of $2\pi$. 
\end{abstract}

\maketitle

\section{Introduction}
The quantum geometric tensor (QGT) \cite{berry89} provides a unified description of the 
local geometry of quantum state manifolds. Its real part defines the quantum metric, 
quantifying the distinguishability of neighboring states, while its imaginary part is related 
to the Berry curvature and geometric phase. The QGT has found applications in condensed 
matter physics \cite{peotta15,picheon16,gianfrate20,marsal24,kang25,yu25}.

For pure states, quantum geometry is uniquely characterized by the Fubini--Study metric 
\cite{provost80} and the Berry curvature \cite{berry84}, which correspond to the real and 
imaginary parts of the QGT, respectively. For mixed states, however, these concepts 
admit different extensions. In particular, mixed-state geometric phases can be defined 
according to different operational principles, leading to distinct notions of quantum geometry. 
An operationally motivated approach is provided by the interferometric geometric phase 
\cite{sjoqvist00} and metric \cite{sjoqvist20}, which connect mixed-state geometry to 
experimentally accessible interference phenomena.

Building on this construction, Ref.~\cite{zhou24} (see also \cite{wang26}) proposed a 
candidate QGT for mixed states, whose real and imaginary parts were interpreted as 
an interferometric quantum metric and a curvature-like quantity, respectively. Here, we 
examine the geometric interpretation of this tensor, focusing on whether its imaginary 
part can be identified with the curvature of a connection on a globally defined U(1) line 
bundle whose holonomy gives a geometric phase.

We note that an alternative QGT has been proposed \cite{wang25} based on the 
Uhlmann--Bures geometry of mixed states \cite{bures69,uhlmann76,uhlmann86}. The 
Uhlmann--Bures construction specifies a global parallel-transport prescription together 
with a connection 1-form \cite{hubner93}. The QGT in Ref.~\cite{wang25} describes a 
geometric structure distinct from the QGT considered here. 
 
\section{The proposed QGT}

Consider a mixed state described by the density operator $\rho$, whose non-zero 
eigenvalues $\lambda_n({\bf R})$ are non-degenerate \cite{remark1,singh03} and 
whose corresponding eigenvectors are $\ket{n({\bf R})}$, with both generally depending 
on some smoothly varying parameters ${\bf R}$. The QGT proposed in \cite{zhou24} 
takes the form (suppressing the ${\bf R}$ dependence for notational convenience)
\begin{eqnarray}
Q_{\mu \nu} = \sum_n \left[ \frac{\partial_\mu \lambda_n \partial_\nu \lambda_n}{4 \lambda_n} +
\lambda_n \bra{\partial_\mu n} \left( \mathbb{1} - \ket{n}\bra{n} \right) \ket{\partial_\nu n}
\right] 
\end{eqnarray}
with $\partial_{\mu} = \partial /\partial R_{\mu}$. The decomposition into real and 
imaginary parts is 
\begin{eqnarray}
    Q_{\mu \nu} =
    g^\text{FR}_{\mu \nu} +
    g^\text{FS}_{\mu \nu} - \frac{i}{2} \Omega_{\mu \nu},
\end{eqnarray}
with $g^\text{FR}_{\mu \nu}$ being the Fisher--Rao information metric, $g^\text{FS}_{\mu\nu}$ 
the weighted sum of Fubini--Study metrics, and
\begin{eqnarray}
\Omega_{\mu \nu} = \sum_n \lambda_n F^{(n)}_{\mu \nu} . 
\end{eqnarray}
Here,  
\begin{eqnarray}
F^{(n)}_{\mu \nu} =
i \left( \bra{\partial_\mu n}\partial_\nu n \rangle - \bra{\partial_\nu n} \partial_\mu n \rangle \right)   
\end{eqnarray}
is the Berry curvature of the $n$-th eigenstate. We introduce the corresponding 
curvature-like 2-form 
\begin{eqnarray}
\Omega = \frac{1}{2}\Omega_{\mu\nu}dR^\mu\wedge dR^\nu = \sum_n\lambda_n F^{(n)},
\label{eq:2form}
\end{eqnarray}
where
\begin{eqnarray}
F^{(n)} = 
\frac{1}{2}F^{(n)}_{\mu\nu} dR^\mu\wedge dR^\nu.
\end{eqnarray}
With this convention, $\Omega$ reduces to the ordinary Berry-curvature 2-form in the 
pure-state limit.

Let $C = \partial S$ be a closed curve spanned by an oriented surface $S$ in parameter 
space. In analogy with pure states, we can write the integral
\begin{eqnarray}
\theta(C)= \int_S \Omega =
\int_S \sum_n \lambda_n F^{(n)} .
\label{eq:weightedbp}
\end{eqnarray}
Assuming the eigenvalues $\lambda_n$ of the density operator remain constant
on $S$ (which they generally do not), this simplifies to 
\begin{eqnarray}
\theta (C) = \sum_n \lambda_n \int_S F^{(n)} = \sum_n \lambda_n \theta^{(n)}_\text{B} (C) ,
\end{eqnarray}
where $\theta^{(n)}_\text{B} (C)$ is the Berry phase of the state $\ket{n}$
undergoing evolution along $C$.

It is useful to contrast the above quantity with the interferometric mixed-state geometric 
phase in unitary evolution introduced in Ref.~\cite{sjoqvist00}. For a cyclic evolution, this 
phase is defined by the argument of a weighted sum of the individual geometric phase factors,
\begin{eqnarray}
\gamma_{\mathrm{int}}(C) = 
\arg\left(\sum_n \lambda_n e^{i\gamma_n(C)}\right),
\end{eqnarray}
rather than by the weighted sum of the phases themselves. Because it depends on the 
phase factors $e^{i\gamma_n (C)}$, each $\gamma_n (C)$ is defined only modulo $2\pi$ 
without producing an ambiguity in $\gamma_{\mathrm{int}} (C)$ \cite{sjoqvist04}. Note 
that the resulting phase $\gamma_{\mathrm{int}} (C)$ itself is defined modulo $2\pi$, as 
appropriate for an interferometric phase. This interferometric construction is therefore 
fundamentally different from $\int_S\Omega$, which is linear in the Berry curvatures. 
Moreover, $\gamma_{\mathrm{int}} (C)$ is not, in general, the holonomy of a globally 
defined U(1) connection on the mixed-state parameter space: the argument of a weighted 
sum of phase factors is non-linear in the individual Berry phases, and a corresponding 
local phase 1-form can even become singular when $\sum_n\lambda_n e^{i\gamma_n (C)}$ 
vanishes. Thus, the interferometric phase provides a gauge-invariant phase observable 
without implying that the weighted Berry-curvature 2-form considered here is the curvature 
of a globally defined U(1) connection. 

We now ask whether the quantity $\theta(C)$ in Eq.~\eqref{eq:weightedbp} can be interpreted 
as a geometric phase associated with a U(1) connection. First, note that the geometric 
phase factor is the holonomy of a connection on the bundle \cite{simon83}: 
\begin{eqnarray}
\text{quantum state space} \hookrightarrow \text{total space}
\overset{\pi}{\to} \text{parameter space},
\nonumber 
\end{eqnarray}
which, for an Abelian connection 1-form $\mathcal{A}$, can be written as
$\exp \left( i \oint_C \mathcal{A}\right)$.
Stokes's theorem gives 
\begin{eqnarray}
\oint_C \mathcal{A} = \int_S \mathcal{F},
\end{eqnarray}
where $C=\partial S$ and $\mathcal{F}=d\mathcal{A}$ is the curvature 2-form.

The first question that should therefore be asked about the proposed new
curvature is whether this relation holds for some connection $1$-form
$\mathcal{A}$. The authors of Ref.~\cite{zhou24} address this
issue, remarking that an explicit solution for the connection may not exist.
One way to check if the proposed quantity $\Omega$ can be interpreted as the
curvature on some manifold is to ensure the Bianchi identity $d\Omega = 0$ \cite{nakahara03}.
Calculating explicitly, we have:
\begin{eqnarray}
d \Omega & = & \sum_n d \lambda_n F^{(n)} + \sum_n \lambda_n d F^{(n)}
\nonumber \\
 & = & \sum_n d \lambda_n F^{(n)},
\end{eqnarray}
where the second equality follows from the pure-state requirement $d F^{(n)} = 0$.
Clearly, for a general mixed state, where the dependence of $\lambda_n$ on the
parameters can vary arbitrarily, $\Omega$ is not a closed form and so cannot be
considered the differential of some connection. Furthermore, considering
again the simplified example in which the $\lambda_n$ remain constant on $S$,
we can write
\begin{eqnarray}
\Omega = \sum_n \lambda_n F^{(n)} = d\left( \sum_n \lambda_n A^{(n)} + \Phi \right) =: dA',
\end{eqnarray}
where $dA^{(n)}=F^{(n)}$ and $\Phi$ is an arbitrary closed 1-form. However, the
eigenstates define line bundles that can have different first Chern numbers
over the parameter manifold. Consequently, their Berry connections need not
combine into a globally defined U(1) connection on the mixed-state parameter 
space. Thus, even when the $\lambda_n$s are constant, the 1-forms $A'$ defined 
locally on charts covering parameter space do not, in general, patch together as a 
connection on a globally defined U(1) line bundle. 

Under the independent gauge transformations
$\ket{n} \rightarrow e^{i\theta_n} \ket{n}$, the Berry connection transforms as 
$A^{(n)} \rightarrow A^{(n)} - d\theta_n$. Consequently,
\begin{eqnarray}
A' \rightarrow A' - 
\sum_n \lambda_n d\theta_n .
\end{eqnarray}
This is not, in general, the transformation law of a single U(1) connection with one 
gauge parameter. Thus, although $\Omega = dA'$ is gauge invariant, $A'$ does not 
generally provide the connection of a conventional U(1) geometric phase for mixed states.

To illustrate that $\Omega$ cannot be treated as a curvature, we consider the simple 
two-level system defined by the Hamiltonian
\begin{eqnarray}
H = \boldsymbol{b} \cdot \boldsymbol{\sigma}
\end{eqnarray}
with $\boldsymbol{b} = b (\sin \theta \cos \phi, \sin \theta \sin \phi, \cos
\theta)$ and $ \boldsymbol{\sigma}$ denotes the vector of Pauli matrices. The eigenvectors 
depend only on the direction $\hat{\boldsymbol b}=\boldsymbol b/b$, so the relevant 
parameter space for the eigenstate geometry is the sphere $S^2$, with the degeneracy 
at $b=0$ excluded. As is well known, the Berry connections of the two eigenstates 
$u_{\pm}$ with energy eigenvalues $\pm b$ of this system are in direct analogy to 
those of the Dirac monopole, with monopole charges $-\frac{1}{2}$ and $+\frac{1}{2}$, 
respectively.

Now consider the mixed state represented by
\begin{eqnarray}
\rho = \lambda_+ \ket{u_+}\bra{u_+} + \lambda_- \ket{u_-}\bra{u_-} 
\end{eqnarray}
with $\lambda_{\pm}$ being constants,  such that $\lambda_+ \neq \lambda_-$ 
and $\lambda_+ + \lambda_- = 1$.
Define the northern and southern hemispheres of the parameter space $S^2$:
\begin{eqnarray}
S_N & := &
\left\{ (\theta,\phi) \middle| 0\leq\theta\leq\frac{\pi}{2} \right\},
\nonumber \\
S_S & := & \left\{ (\theta,\phi) \middle| \frac{\pi}{2}\leq\theta\leq\pi \right\},
\end{eqnarray}
so that
\begin{eqnarray}
S^2=S_N\cup S_S, \qquad \partial S_N=C=-\partial S_S,
\end{eqnarray}
where $C$ is the equator traversed counter-clockwise as seen from the north pole 
of the parameter sphere. The Berry connections $A_\pm = i \langle u_\pm|du_\pm\rangle$ 
are each defined locally on two charts of the sphere, overlapping at the equator: $U_N$, 
which includes the north pole, and $U_S$, which includes the south pole. On the overlap, 
their gauge transformations are characterized by the Dirac quantization condition 
\begin{eqnarray}
A_\pm^N=A_\pm^S\mp d\phi ,
\end{eqnarray}
where the signs correspond to the convention
\begin{eqnarray}
\int_{S^2}F_\pm=\mp 2\pi.
\label{eq:convention}
\end{eqnarray}
 
 Since $\lambda_\pm$ are constant, the integral of the proposed curvature-like 2-form 
 $\Omega$ over a surface with boundary $C$ can be evaluated using either of the local 
 Berry connections. Using the northern chart gives
 \begin{eqnarray}
 \alpha_N
 &=&
 \oint_C
 \left(
 \lambda_+ A_+^N+\lambda_- A_-^N
 \right)
 \nonumber \\
  & = & \lambda_+\int_{S_N}F_+ +
 \lambda_-\int_{S_N}F_- ,
 \end{eqnarray}
 whereas using the southern chart gives
 \begin{eqnarray}
 \alpha_S
  & = & \oint_C \left(
 \lambda_+ A_+^S + \lambda_- A_-^S
 \right)
 \nonumber \\
  & = & -\lambda_+\int_{S_S}F_+
 -\lambda_-\int_{S_S}F_- .
 \label{eq:geometricalphase}
 \end{eqnarray}
 The minus signs in the second expression follow from $\partial S_S=-C$.
 
 The difference between the two evaluations is therefore
 \begin{eqnarray}
 \alpha_N-\alpha_S
  & = & \lambda_+\int_{S^2}F_+ + 
 \lambda_-\int_{S^2}F_-
 \nonumber \\ 
  & = & 2\pi(\lambda_--\lambda_+) , 
 \label{eq:surfaceambiguity}
 \end{eqnarray}
 where we have used Eq.~\eqref{eq:convention}. Thus, the two surfaces bounded by 
 the same closed curve $C$ give values that differ by $2\pi(\lambda_--\lambda_+)$. 
 For a full-rank ($0<\lambda_\pm<1$), non-degenerate ($\lambda_+\neq\lambda_-$) 
 mixed state, we have 
 \begin{eqnarray}
 0<|\lambda_--\lambda_+|<1.
 \end{eqnarray}
Consequently, although $\Omega$ is a globally defined 2-form, its surface integral 
$\alpha = \int_S\Omega$ is not determined by the boundary curve $C$ modulo $2\pi$ 
for a generic full-rank, non-degenerate mixed state. Thus, $\Omega$ cannot be 
interpreted as the curvature of a U(1) connection whose holonomy defines a geometric 
phase associated solely with the closed path $C$.
 
 In the pure-state limit, for example $\lambda_+ = 1$ and $\lambda_- = 0$, 
 Eq.~\eqref{eq:surfaceambiguity} gives
 \begin{eqnarray}
 \alpha_N-\alpha_S=-2\pi,
 \end{eqnarray}
 which is an allowed $2\pi$ ambiguity of the ordinary Berry phase. Thus, the pure-state 
 limit is recovered consistently, whereas the ambiguity becomes non-quantized for a 
 full-rank, non-degenerate mixed state.

The above analysis can be recast in terms of first Chern numbers. For the Hamiltonian 
$H = \mathbf{b} \cdot \boldsymbol{\sigma}$, the two eigenstate line bundles have first 
Chern numbers  
\begin{eqnarray}
-\mathscr{C}_+ = \mathscr{C}_- = 1 .
\end{eqnarray}
If one attempts to define an analogous Chern number for the mixed state by integrating 
the proposed curvature $\Omega$, one obtains 
\begin{eqnarray}
\frac{1}{2\pi} \int_{S^2} \Omega = \lambda_+ \mathscr{C}_+ + \lambda_- \mathscr{C}_- =  
\lambda_- - \lambda_+ ,
\end{eqnarray}
which, as noted above, is generally not an integer. This is incompatible with the defining 
property of a first Chern number, which is an integer-valued topological invariant associated 
with a complex line bundle \cite{nakahara03}. More fundamentally, as demonstrated above, 
the corresponding "holonomy factor" is surface-dependent, so $\Omega$ does not define 
a globally consistent curvature for the mixed state in the sense required for a Chern class. 
Thus, the weighted quantity $\lambda_+ \mathscr{C}_+ + \lambda_- \mathscr{C}_-$ should 
not be interpreted as a first Chern number: for a full-rank, non-degenerate mixed state, 
a first Chern number cannot be defined from $\Omega$. 

\section{Conclusions}
We have examined the geometric interpretation of the quantum geometric tensor proposed 
in Ref.~\cite{zhou24} for mixed states. While its real part decomposes into Fisher--Rao 
and weighted Fubini--Study contributions, its imaginary part is proportional to a weighted 
sum of the eigenstate Berry curvatures. The two-level example shows that the constituent 
eigenstates can carry different first Chern numbers, which implies that the corresponding 
weighted Berry-phase integral over different surfaces bounded by the same closed path 
can differ by $2\pi(\lambda_- -\lambda_+)$, which is not an integer multiple of $2\pi$ for 
full-rank, non-degenerate states. Thus, although the proposed 2-form is gauge invariant, 
its surface integral cannot generally be interpreted as a physical mixed-state geometric 
phase. This highlights the importance of gauge invariance, global consistency, and a 
well-defined holonomy when assigning geometric meaning to mixed-state generalizations 
of the QGT.

\section*{Acknowledgements}
E. S. acknowledges financial support from the Swedish Research Council (VR) 
through Grant No. 2025-05249.


\begin{thebibliography}{99}
\bibitem{remark1} The mixed-state geometric phase \cite{sjoqvist00} and quantum 
metric \cite{sjoqvist20} are well-defined only for non-degenerate density operators, 
for which the eigenvectors can be assigned uniquely up to independent U(1) gauge 
transformations. If the density operator becomes degenerate, the corresponding 
eigenspaces acquire a non-Abelian gauge freedom \cite{singh03}, in contrast to the 
Abelian structure considered in the present work.
\end{thebibliography}
\end{document}